\documentclass{iopart}
\usepackage{setstack}
\usepackage{graphics}
\usepackage{graphicx}
\usepackage{amssymb}
\usepackage{amsopn}
\usepackage[]{hyperref}
\usepackage[]{color}
\usepackage[all]{xy}
\newcommand{\ff}[2]{\frac{#1}{#2}}
\newcommand{\dd}[0]{\mathrm{d}}
\newcommand{\bb}[0]{\begin{eqnarray}}
\newcommand{\ee}[0]{\end{eqnarray}}
\newcommand{\nn}{\nonumber}

\newcommand{\four}[1]{\mathcal{F}\left[#1\right]}

\newcommand{\set}[1]{\mathfrak{X}_{#1}}
\newcommand{\moy}[1]{\langle #1 \rangle}
\newcommand{\MAX}[2]{\mathop{\mathrm{Max}^{(#1)}}_{#2}}

\newcommand{\eqva}[2]{\mathop{\lsim}\limits_{\; \; #1 \rightarrow #2}}

\begin{document}
\title{Generalised extreme value statistics and sum of correlated variables}

\author{Eric Bertin$^{\dag}$ and  Maxime Clusel$^\ddag$}
\address{\dag\ Department of Theoretical Physics, University of Geneva, 
CH-1211 
Geneva 4, Switzerland}
\address{\ddag\ Institut Laue-Langevin, 6 rue Horowitz BP156 X, 38042 Grenoble
cedex, France}

\ead{eric.bertin@physics.unige.ch, clusel@ill.fr}

%---- ABSTRACT ----%
\begin{abstract}
We show that generalised extreme value statistics --the statistics
of the $k^\mathrm{th}$ largest value among a large set of random variables--
can be mapped onto a problem of random sums.
This allows us to identify classes of non-identical
and (generally) correlated random variables with a sum distributed according
to one of the three ($k$-dependent) asymptotic distributions of extreme
value statistics, namely the Gumbel, Fr\'echet and Weibull distributions.
These classes, as well as the limit distributions, are naturally extended
to real values of $k$, thus providing a clear interpretation to the onset of
Gumbel distributions with non-integer index $k$ in the statistics of global
observables. This is one of the very few known generalisations of the
central limit theorem to non-independent random variables.
Finally, in the context of a simple physical model, we relate the index $k$
to the ratio of the correlation length to the system size,
which remains finite in strongly correlated systems.

\end{abstract}
\pacs{\\02.50.-r (Probability theory, stochastic processes, and statistics);\\05.40.-a (Fluctuation phenomena, random processes, noise, and Brownian motion); }
\submitto{\JPA}

%\maketitle
%\tableofcontents
%\hrule

%---- INTRODUCTION ----%
\section{Introduction}

One of the cornerstones of probability theory is the celebrated central limit
theorem, stating that under general assumptions, the distribution of the sum of independent
random variables converges, once suitably rescaled, to a Gaussian
distribution. This theorem provides one of
the foundations of statistical thermodynamics and, from a more practical
point of view, legitimates the use of Gaussian
distributions to describe fluctuations appearing in experimental or
numerical data.\\
However, this general theorem breaks down in several situations of physical
interest, when the assumptions made to derive it are no longer fulfilled.
For instance, if the random variables contributing to the sum have an
infinite variance, the distribution of the sum is no longer Gaussian,
but becomes a L\'evy distribution \cite{Levy54}. This has deep physical
consequences, for instance in the context of laser cooling \cite{Bardou02}
or in that of glasses \cite{BG90,Bouchaud92}. Though the central limit
theorem applies to a class of correlated variables, the martingale differences \cite{billingsley,ibragimov}, it is generally inapplicable to sums of strongly correlated, or alternatively, strongly non-identical random variables.
%Another way to break the usual form of the central limit theorem is
%to consider sums of strongly correlated variables, or alternatively,
%sums of strongly non-identical random variables. 
In this case, no general mathematical theory is available,
but experimental as well as numerical distributions of the fluctuations of
global observables in correlated systems often present an asymmetric
shape, with an exponential tail on one side and a rapid fall-off on the
other side. Such distributions are usually well described by the
so-called generalised Gumbel distribution with a real index $a$. The latter
is one of the limit distributions appearing, for integer values $a=k$,
as the distribution of the $k^\mathrm{th}$ extremal value of a set of independent and identically distributed random variables \cite{gumbel,galambos}.
For instance, Bramwell and co-workers
argued that the generalised Gumbel distribution with
$a\approx 1.5$ reasonably describes the large scale fluctuations in
many correlated systems \cite{bramwell00}. Since then, the generalised
Gumbel distribution has been reported in various contexts
\cite{noullez02,chamon04,pennetta04,duri05,varotsos05}. Yet, the
interpretation of such a distribution in the context of
extreme value statistics (EVS) is
far from obvious. This issue led to a debate around the possible existence
of a (hidden) extreme process which might dominate the statistics
of sums of correlated variables in physically relevant situations
\cite{dahlstedt01,watkins02,bramwell02,clusel04}. Still, no evidence for
such a process has been found yet. And even if one accepts that some
extremal process is at play, a conceptual difficulty remains,
as speaking of the $k^\mathrm{th}$ largest value for non-integer $k$
simply does not make sense.\\
In this article we propose another interpretation of the usual EVS in
terms of statistics of sums of correlated variables. Within this framework
the extension to real positive values of $k$ is straightforward and avoids
any logical problem. Usual EVS then appears as a particular case of a more
general problem of limit theorem for sums of correlated variables
belonging to a given class.\\
The article is organised as follows. In Section 2 we present the link
between usual EVS and sums of correlated variables, and propose a
more general problem of random sums, which is studied is the next sections.
In Section 3, we present in detail the simple problem of sum
--already presented shortly by one of us in a previous publication
\cite{bertin05}-- associated with the EVS of exponential random variables.
Section 4 deals with the extension to the more general problem of random
sums defined in Section 2, and it is shown that the asymptotic distribution
of the sum, is either a (generalised) Gumbel, Fr\'echet or Weibull
distribution. Finally, we discuss in Section 5 some physical interpretations
of our results.

\section{Equivalence between extreme value statistics and sums of correlated
variables}

\subsection{Standard extreme value statistics}

The asymptotic theory of EVS, which found applications for instance in physics \cite{Bouchaud97}, hydrology \cite{Katz02}, seismology \cite{Sornette96} and finance \cite{Longin00}, is also part of the important results
of probability theory. Consider a random variable $X$,
distributed according to the probability density $P$,
and let $\set{N}$ be a set of $N$ realisations of this random variable.
Now instead of considering the sum of these $N$ realisations as in the central limit theorem, one is interested in the
study of the distribution of the $k^\mathrm{th}$ maximal value $z_k$ of
$\set{N}$. The major result of EVS is that,
in the limit $N \rightarrow \infty$,
the probability density of $z_k$ does not depend on the details of the
original distribution $P$: There are only three different types of
distributions, depending on the asymptotic behaviour of $P$
\cite{gumbel,galambos}. In particular, in the case where $P$ decreases
faster than any power laws at large $x$, the variable $z_k$ is distributed
according to the Gumbel distribution
\bb \label{nGk}
G_k(z_k)=\ff{k^k \theta_k}{\Gamma(k)} \exp \left[-k\theta_k(z_k+\nu_k)-
ke^{-\theta_k(z_k+\nu_k)}\right],\\
\nn \theta_k^2=\Psi'(k),\quad
\nu_k=\ff{1}{\theta_k}\big[\ln k - \Psi(k)\big],
\ee
where $\Psi(x)=\ff{\dd}{\dd x} \ln \Gamma(x)$ is the digamma function.
Note that from the point of view of EVS, $k\in \mathbb{N}^*$ by construction.
Yet, if one forgets about EVS, the expression (\ref{nGk}) is also formally
valid for $k=a$ real and positive.

\subsection{Reformulation as a problem of sum}
Let us consider again the set $\set{N}$ defined here above.
We then introduce the ordered set of random variables, $z_k=x_{\sigma(k)}$,
where $\sigma$ is the ordering permutation such that $z_1\ge z_2 ... \ge z_N$.
One can now define the increments of $z$ with the following relations:
\bb \nn
u_k &\equiv& z_k-z_{k+1},\ \forall \ 1\le k \le N-1,\\ \nn
u_N &\equiv& z_N.
\ee
Then by definition, the following identity holds
\bb \label{reffond}
z_n \equiv \MAX{n}{}\Big[ \set{N} \Big]= \sum_{k=n}^N u_k,
\ee
where $\MAX{n}{}$ is the $n^\mathrm{th}$ largest value of the set $\set{N}$,
showing that an extreme value problem may be reformulated as a problem
of sum of random variables $U_k$.\\
Besides, as these new random variables have been obtained through an ordering process, they are
\textit{a priori} non-independent and non-identically distributed. This is
confirmed by the computation of the joint probability
$\tilde{J}_{k,N}(u_k,...,u_N)$ which is given by
\bb \fl
\tilde{J}_{k,N}(u_k,...,u_N)&=&N!\int_0^{\infty} \dd z_N P(z_N)
\int_{z_N}^{\infty} \dd z_{N-1} P(z_{N-1}) ...\int_{z_2}^{\infty}
\dd z_1 P(z_1) \nn \\
&& \qquad \qquad \times
\delta(u_N-z_N) \prod_{n=k}^{N-1}\delta(u_n-z_n+z_{n+1}),
\nn \\
&=& N! \left[\prod_{n=k}^{N}P\left(\sum_{i=n}^N u_i\right)\right]
\int_{u_k+...+u_N}^{\infty}\dd z_{k-1} P(z_{k-1})...\int_{z_2}^{\infty}\dd
z_{1} P(z_{1}).\nn
\ee
By recurrence it is easy to demonstrate the following relation,
\bb \nn
\int_{u}^{\infty}\dd z_{k-1} P(z_{k-1})...\int_{z_2}^{\infty}\dd z_{1}
P(z_{1})=\ff{1}{(k-1)!}F(u)^{k-1},
\ee
with $F(u)\equiv \int_u^{\infty}\dd z  \ P(z)$, leading to 
\bb \nn
\tilde{J}_{k,N}(u_k,...,u_N)=\ff{N!}{(k-1)!}\; F\left(\sum_{i=k}^N u_i\right)^{k-1}
\prod_{n=k}^{N}P\left(\sum_{i=n}^Nu_i\right).
\ee
The probability $\tilde{J}_{k,N}$ is actually a probability at $N'=N+1-k$ points,
and a shift of indices allows one to write it in the final form:
\bb \fl \label{JN}
\qquad \qquad J_{N'}(u_1,...,u_{N'})=\ff{(N'+k-1)!}{(k-1)!}\;
F\left(\sum_{i=1}^{N'}u_i\right)^{k-1}
\prod_{n=1}^{N'}P\left(\sum_{i=n}^{N'} u_i\right).
\ee
In the general case, this expression does not factorise, \textit{i.e.} one
cannot find $N'$ functions $\pi_n$ such that
\bb \nn \label{Jfact}
J_{N'}(u_1,...,u_{N'})=\prod_{n=1}^{N'}\pi_n(u_n),
\ee
and the random variables $U_n$ are thus non-independent.

\subsection{Extension of the joint probability}
Up to now, we only reformulated the usual extreme value statistics as a
problem of sum, by some trivial manipulations. Basically, we have two
equivalent problems with the same asymptotic distribution, e.g., the
standard Gumbel distribution if $P$ is in the Gumbel class.
Let us now point out that Eq.~(\ref{JN}) is a particular case of the
following joint distribution
\bb \label{JNgen}
J_{N}(u_1,...,u_{N}) = \ff{\Gamma(N)}{Z_N}\;
\Omega\left[ F\left(\sum_{n=1}^{N}u_n\right)\right]
\prod_{n=1}^{N}P\left(\sum_{i=n}^{N} u_i\right)
\ee
where $\Omega(F)$ is an (arbitrary) positive function of $F$, and where
$Z_N$ is given by
\bb \label{eq-ZN}
Z_N = \int_0^1 \dd v\, \Omega(v)\, (1-v)^{N-1}.
\ee
Eq.~(\ref{JN}) is recovered by choosing $\Omega(F)=F^{k-1}$.
Note that by extending the definition of the joint probability, we lost the
equivalence between the two original problems. In other words, we are
generalising the problem of statistics of sums and \textit{not} the
extreme value one. In the rest of this article, we study the limit
distribution of a sum of correlated random variables satisfying the
generalised joint probability (\ref{JNgen}).\\
Let us note finally that one can easily generate from the joint probability
(\ref{JNgen}) other joint probabilities that lead to the same asymptotic
distribution for the sum, by summing over a set $\mathcal{S}$ of permutations
over $[1,...,N]$
\bb \nn
J_{N}^{\mathcal{S}}(u_1,...,u_{N}) = \ff{1}{\mathcal{N}(\mathcal{S})}
\sum_{\sigma \in \mathcal{S}} J_{N}(u_{\sigma(1)},...,u_{\sigma(N)}),
\ee
$\mathcal{N}(\mathcal{S})$ being the cardinal of S.
This may allow in particular some symmetry
properties between the variables to be restored.
Indeed, starting for instance from independent and non-identically
distributed random variables, this procedure leads to a set of
non-independent and identically distributed random variables with the same 
statistics for the sum.

\section{The exponential case}
Though in general the joint probability does not factorise, there is a
particular case where it does.
For pedagogical purposes, we discuss this case separately, before dealing
with the general case in Section~\ref{sect-gen}.
To allow the joint probability $J_{N}$ defined by (\ref{JNgen}) to factorise, the function $P$ has to satisfy the following property:
\bb \nn
\forall (x,y),\ P(x+y)=P(x)P(y).
\ee
That is to say that $P$ must be exponential: $P(x)=\kappa e^{-\kappa x}$,
with $\kappa>0$.
As a result, one has $F(x)=e^{-\kappa x}$, and the factorisation criterion
on $J_{N}$ translates into the following condition on $\Omega$
\bb \nn
\forall (F_1,F_2),\ \Omega(F_1 F_2)=\Omega(F_1)\Omega(F_2),
\ee
so that $\Omega$ must be a power law: $\Omega(F) = F^{a-1}$ (the prefactor
may be set to $1$ without loss of generality),
with $a>0$ to ensure the convergence of the integral defining $Z_N$
given in Eq.~(\ref{eq-ZN}).
In such a case, one finds
\bb
Z_N = \ff{\Gamma(N)\Gamma(a)}{\Gamma(N+a)},
\ee
and the joint probability could be written in the factorised form
(\ref{Jfact}),
with 
\bb \label{pin}
\pi_n(u_n)=(n+a-1)\kappa \, e^{-(n+a-1)\kappa u_n}.
\ee
Therefore, for $a=k$ integer, if one sums independent and non-identically
distributed random variables $U_n$, $n=1,...,N$, obeying (\ref{pin}),
the distribution of the sum converges to a Gumbel distribution $G_k$
in the limit $N \rightarrow \infty$.
Note that the particular case $k=1$ has been previously studied by Antal
\etal in the context of $1/f$ noise \cite{antal01}.
In the following of this section, we shall establish that for $a$ real,
the distribution of the sum of the $U_n$'s converges towards
the generalised Gumbel distribution $G_a$ for $N \rightarrow \infty$, as
announced in a previous publication \cite{bertin05}.\\
To this aim, let us define the random sum
$S_N=\sum_{n=1}^N U_n$, where $U_n$ is distributed according to (\ref{pin}),
with $a$ real and positive. The distribution
of $S_N$ is denoted by $\Upsilon_N$. As the $U_n$'s are independent,
the Fourier transform of $\Upsilon_N$ is simply given by
\bb \label{TFPN}
\four{\Upsilon_N}(\omega)=\prod_{n=1}^N \four{\pi_n}(\omega)=\prod_{n=1}^N
\left( 1+\ff{i\omega}{\kappa(n+a-1)} \right)^{-1}.
\ee
The first two moments of the distribution read
\bb \label{moy}
\moy{S_N} &=& \sum_{n=1}^{N} \moy{U_n} =\ff{1}{\kappa} \sum_{n=1}^N
\ff{1}{n+a-1}, \\ \label{sigma}
\sigma_N^2 &=& \sum_{n=1}^{N} \mathrm{Var}(U_n)=\ff{1}{\kappa^2}\sum_{n=1}^N
\ff{1}{(n+a-1)^2}.
\ee
In order to get a well-defined distribution in the limit $N \rightarrow
\infty$, let us introduce the reduced variable $\mu$ by
\bb \nn 
\mu=\ff{s-\moy{S_N}}{\sigma},\ \mathrm{with} \ \sigma = \lim _{N\rightarrow
\infty} \sigma_N.
\ee
Note that the fact that $\sigma< \infty$ breaks the Lindeberg's condition, allowing an eventual breakdown of the central limit theorem \cite{feller1}.
The distribution of $\mu$, $\Phi_N$, is then given by
$\Phi_N(\mu)=\sigma \Upsilon_N(\sigma \mu + \moy{S_N})$.
The Fourier transform of $\Phi=\lim_{N\rightarrow \infty}
\Phi_N$ reads, using (\ref{TFPN},\ref{moy})
\bb \fl \nn
\four{\Phi_\infty}(\omega)=\lim_{N\rightarrow \infty} \four{\Phi_N}(\omega)=
\prod_{n=1}^{\infty} \left(1+\ff{i\omega}{\sigma \kappa (n+a-1)}\right)^{-1}
\exp \left(\ff{i\omega}{\sigma \kappa(n+a-1)}\right).
\ee
The last part of the previous expression has to be compared with the Fourier
transform given in Appendix A, leading to 
\bb \nn
\Phi_\infty(\mu)=G_{a}(\mu).
\ee
This result shows that it is possible to obtain quite directly the
generalised Gumbel distribution with a real index $a$,
not from an extremal process, but from a
sum of independent non-identically distributed random variables. As already
pointed out, the fact that the random variables $U_n$ are uncorrelated
is a specificity of the exponential distribution. A natural question is 
then to know whether this result survives for a more general class of
distributions $P$ and functions $\Omega$.
We address this question in the next section. 

\section{Generalised extreme value distributions} \label{sect-gen}
In the general case, the random variables $U_n$ are non-independent,
so we have to deal with the joint probability (\ref{JNgen}).
Accordingly, the route to the asymptotic limit distribution will be
quite different from the exponential case.

\subsection{Distribution of the sum for finite $N$}
Consider a set of realisations $\{u_n\}$ of $N$ (correlated) random variables
$U_n$, with the joint probability (\ref{JNgen}).
We then define as above the random variable $S_N=\sum_{n=1}^N U_n$, and let
$\Upsilon_N$ be the probability density of $S_N$.
Then $\Upsilon_N$ is given by
\bb \nn
\Upsilon_N(s)=\int_0^{\infty}\dd u_N ... \dd u_1 J_N(u_1,...,u_N)\; \delta
\left(s-\sum_{n=1}^N u_n \right).
\ee
Inserting (\ref{JNgen}), one obtains
\bb \nn
\Upsilon_N(s)=\ff{\Gamma(N)}{Z_N}\, P(s)\, \Omega\big(F(s)\big)\, I_N(s),
\ee
with
\bb \fl \nn
I_N(s)=\int_{0}^{\infty} \dd u_N P(u_N) \int_{0}^{\infty} \dd u_{N-1}
P(u_N+u_{N-1})...\int_{0}^{\infty} \dd u_1 \delta\left(s-\sum_{n=1}^N u_n
\right).
\ee
To evaluate $I_N$, let us start by integrating over $u_1$, using
\bb \nn
\int_{0}^{\infty} \dd u_1 \delta\left(s-\sum_{n=1}^N u_n \right)=\Theta
\left(s-\sum_{n=2}^N u_n \right),
\ee
where $\Theta$ is the Heaviside distribution. This changes the upper bound
of the integral over $u_2$ by $u_2^\mathrm{max}=\max \left( 0, s-\sum_{n=2}^N
u_n \right)$. Then the integration over $u_2$ leads to
\bb \nn
\int_0^{u_2^\mathrm{max}} \dd u_2 P(u_2)=\left[ F\left( \sum_{n=3}^N u_n
\right)-F(s)\right] \Theta\left(s-\sum_{n=3}^N u_n \right),
\ee
By recurrence it is then possible to show that
\bb \nn
I_N(s)=\ff{1}{\Gamma(N)}\Big(1-F(s)\Big)^{N-1},
\ee
finally yielding the following expression for $\Upsilon_N$:
\bb \label{PsiN}
\Upsilon_N(s)=\ff{1}{Z_N}\,P(s)\,\Omega\big(F(s)\big)\,
\Big(1-F(s)\Big)^{N-1}. 
\ee
In the following sections, we assume that $\Omega(F)$ behaves
asymptotically as a power law $\Omega(F) \sim \Omega_0 \, F^{a-1}$
when $F \to 0$ ($a>0$).
Under this assumption, we deduce from Eq.~(\ref{PsiN}) the different
limit distributions associated with the different classes of asymptotic
behaviours of $P$ at large $x$.

\subsection{The Gumbel class}
In this section we focus on the case where $P$ is in the Gumbel class,
that is, $P(x)$ decays faster than any power law at large $x$.
Note that the exponential case studied above precisely belongs to this class.
Our aim is to show that, after a suitable rescaling of the variable, the limit
distribution obtained from (\ref{PsiN}) is the generalised Gumbel
distribution $G_a$, where $a$ characterises the asymptotic behaviour of
$\Omega(F)$ for $F \rightarrow 0$.

\begin{figure}
\includegraphics[scale=0.5]{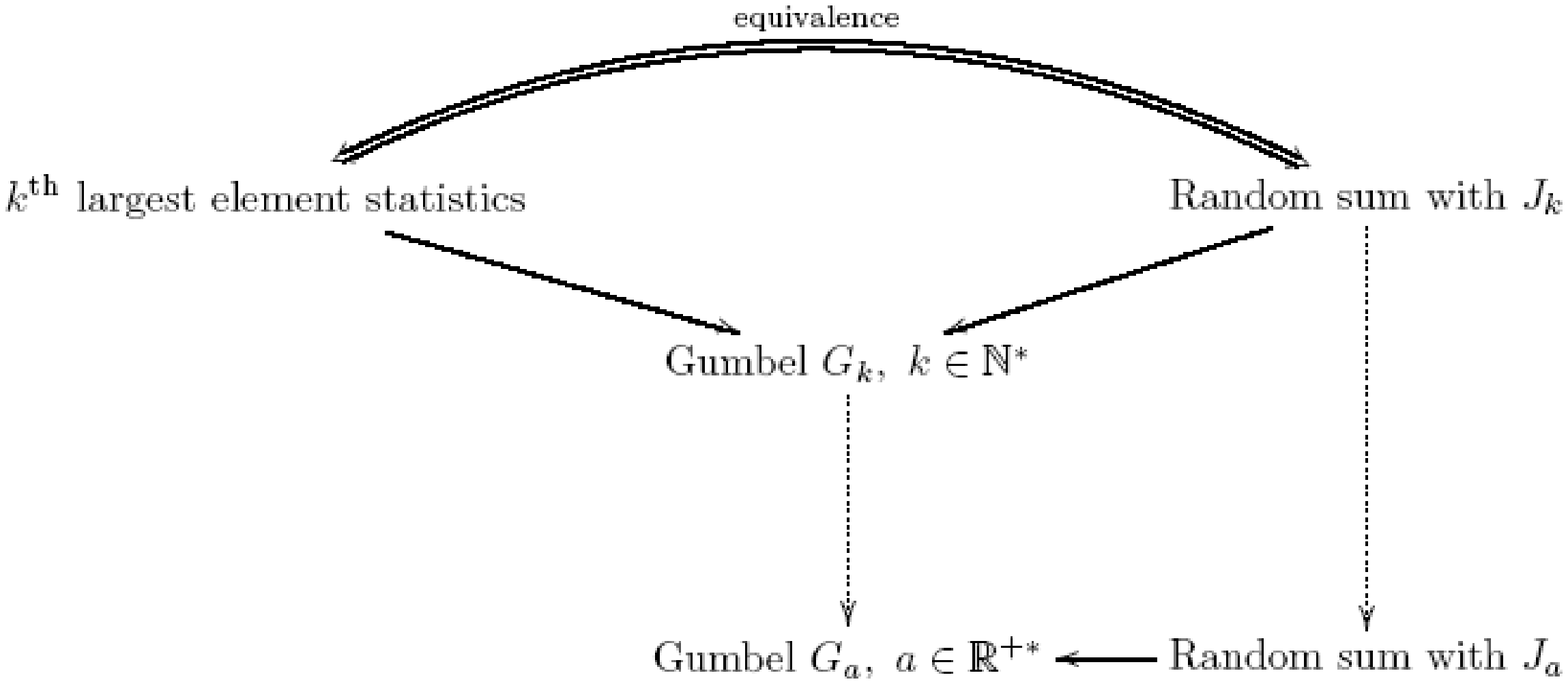}
% \xymatrix{
%  k^{\mathrm{th}}\ \mathrm{largest\ element\ statistics} \ar@/^3pc/
% @{<=>}[rr]^{\mathrm{equivalence}} \ar@//[dr] &  &\mathrm{Random\ sum\
% with\ }J_k   \ar@{.>}[ddd] \ar@//[dl]     \\
%  &\mathrm{Gumbel\ } G_k,\ k\in \mathbb{N}^* \ar@{.>}[dd]  & \\
%  & & \\
%   & \mathrm{Gumbel\ } G_a,\ a\in \mathbb{R}^{+*}  & \mathrm{Random\ sum\
% with\ } J_a \ar@//[l]
% }  
\caption{\label{schema} Generalisation of EVS \textit{via} sums of 
non-independent variables.}
\end{figure}

To that purpose, we define $s^*_N$ by $F(s^*_N)=a/N$. If $a$ is an
integer, this is nothing but the typical value of the $a^\mathrm{th}$
largest value of $s$ in a sample of size $N$. As $P$ is unbounded we have 
\bb \nn
\lim_{N\rightarrow \infty} s^*_N=+\infty.
\ee
Let us introduce $g(s)=-\ln F(s)$ and, assuming $g'(s_N) \ne 0$,
define the rescaled variable $v$ by
\bb \label{def-v}
s=s^*_N+\ff{v}{g'(s^*_N)}.
\ee
For large $N$, one can perform a series expansion of $g$ around $s^*_N$:
\bb \label{expansion} \nn
g(s)=g(s^*_N)+v+\sum_{n=1}^{\infty} \ff{1}{n!}
\ff{g^{(n)}(s^*_N)}{g'(s^*_N)^n}\, v^n.
\ee
For $P$ in the Gumbel class, $g^{(n)}(s^*_N)/g'(s^*_N)^n$ is bounded as
a function of $n$ so that the series converges. In addition, one has
\bb \nn
\lim_{N \rightarrow \infty} \ff{g^{(n)}(s^*_N)}{g'(s^*_N)^n} =0,\ \forall
n\ge2,
\ee
so that $g(s)$ may be written as
\bb \label{gN}
g(s) = g(s^*_N)+v+\varepsilon_N(v), \quad \mathrm{with}
\lim_{N \rightarrow \infty} \varepsilon_N(v) = 0.
\ee
Given that $P(s)=g'(s)\,F(s)$, one gets using Eqs.~(\ref{PsiN}) and
(\ref{gN})
\bb \nn
\Phi_N(v) &=& \ff{1}{g'(s_N^*)} \Upsilon_N(s)\\ \nn
&=& \ff{1}{Z_N} \ff{g'(s)}{g'(s_N^*)}\, F(s)\, \Omega\big(F(s)\big)
\Big( 1-F(s) \Big)^{N-1},
\ee
where $s$ is given by Eq.~(\ref{def-v}). For $P$ in the Gumbel class,
it can be checked that, for fixed $v$
\bb \nn
\lim_{N \rightarrow \infty} \ff{g'\big(s_N^*+v/g'(s_N^*)\big)}{g'(s_N^*)}=1.
\ee
Besides, $F\big(s_N^*+v/g'(s_N^*)\big) \rightarrow 0$ when
$N \rightarrow \infty$, so that one can use the small $F$ expansion of
$\Omega(F)$. Altogether, one finds
\bb \nn
\Phi_N(v) \sim \ff{\Omega_0}{Z_N} \left(\ff{a}{N}\right)^a
e^{-av-a\varepsilon_N(v)} \left[1-\ff{a}{N}\, e^{-v-\varepsilon_N(v)}
\right]^{N-1}
\ee
Using a simple change of variable in Eq.~(\ref{eq-ZN}), one can show that
\bb \nn
\lim_{N \rightarrow \infty} \ff{N^a Z_N}{\Omega_0} = \Gamma(a)
\ee
It is then straightforward to take the asymptotic limit $N \to \infty$,
leading to
\bb \nn
\Phi_{\infty}(v)=\ff{a^a}{\Gamma(a)}\exp\left[-av-ae^{-v} \right].
\ee
In order to recover the usual expression for the generalised Gumbel
distribution, one simply needs to introduce the reduced variable
\bb \nn
\mu=\ff{v-\moy{v}}{\sigma_v},
\ee
with, $\Psi$ being the digamma function,
\bb
\nn \moy{v}=\ln a - \Psi(a),\quad \sigma_v^2=\Psi'(a).
\ee
The variable $\mu$ is then distributed according to a generalised Gumbel
distribution (\ref{nGk}).\\
To sum up, if one considers the sum $S_N$ of $N \gg 1$ random variables
linked by the joint probability (\ref{JNgen}), then the asymptotic
distribution of the reduced variable $\mu$ defined by
\bb \nn
\mu=\ff{s_N-\moy{S_N}}{\sigma_N},
\ee
with
\bb
\nn \moy{S_N}=s^*_N+\ff{\ln a - \Psi(a)}{g^{'}(s^*_N)},\quad
\sigma_N = \ff{\sqrt{\Psi'(a)}}{g^{'}(s^*_N)}.
\ee
is the generalised Gumbel distribution (\ref{nGk}).
More generally, the approach developed in this paper to relate
EVS with sums of non-independent random variables, and then generalise
the problem of sum, may be summarised as shown on Fig.~\ref{schema}
on the example of the Gumbel class.

\subsection{Extension to the Fr\'echet and Weibull classes}
Let us now consider the cases where $P$ belongs either to the
Fr\'echet class, that is $P$ has a power law tail
\bb \nn
P(x) \eqva{x}{\infty} \ff{P_0}{x^{1+\eta}},
\quad \eta>0,
\ee
with $P_0>0$ a constant, or to the Weibull class, that is $P$ has an upper
bound $A$ and behaves as a power law in the vicinity of $A$
\bb \nn
\forall x>A,\ P(x) = 0 \nn \\
P(x) \eqva{x}{A^-} \tilde{P}_0\, (A-x)^{\beta-1}, \quad \beta>0.
\ee
with $\tilde{P}_0$ an arbitrary positive prefactor.
A calculation similar to the one used in the Gumbel case allows the
asymptotic distributions to be determined.
Considering first the Fr\'echet class, we define $s_N^*$ through
$F(s_N^*)=a/N$ as above, and introduce the scaling variable $v=s/s_N^*$.
This gives for the distribution $\Phi_N^{\mathrm{f}}(v)$
\bb \nn
\Phi_N^{\mathrm{f}}(v) \sim \ff{ a^a \Omega_0}{N^a Z_N} \ff{\eta}{v^{1+a\eta}}
\left( 1-\ff{a}{Nv^{\eta}} \right)^{N-1}.
\ee
Taking the limit $N \rightarrow \infty$, the asymptotic distribution reads
\bb \nn
\Phi_{\infty}^{\mathrm{f}}(v) = \ff{a^a}{\Gamma(a)}\ff{\eta}{v^{1+a\eta}}
e^{-a/v^{\eta}}.
\ee
One can rescale the variable $v$ with respect either to the mean value or
to the standard deviation $\sigma$. In this latter case, one finds
\bb \nn
\sigma=\ff{a^{1/\eta}}{\Gamma(a)}\left[\Gamma(a)\Gamma\left(a-\ff{2}{\eta}
\right)-\Gamma\left(a-\ff{1}{\eta}\right)^2\right]^{1/2}.
\ee
Introducing $\mu=v/\sigma$ we get
\bb \nn
F_a(\mu)&=&\ff{\eta \Lambda^a}{\Gamma(a)}\ff{1}{\mu^{1+a\eta}}\exp\left(
-\Lambda \mu^{-\eta}\right),\\ \nn
\Lambda&=&\Gamma(a)^\eta \left[\Gamma(a)\Gamma\left(a-\ff{2}{\eta}\right)
-\Gamma\left(a-\ff{1}{\eta}\right)^2\right]^{-\eta/2}.
\ee
Calculations for the Weibull class are very similar, given that the
scaling variable is now $v=(A-s)/(A-s_N^*)$. Skipping the details,
the asymptotic distribution reads
\bb \nn
\Phi_{\infty}^{\mathrm{w}}(v) = \ff{\beta\, a^a}{\Gamma(a)}\,v^{a\beta-1}
e^{-a v^{\beta}}
\ee
After rescaling to normalise the second moment 
\bb \nn
\sigma=\ff{a^{-1/\beta}}{\Gamma(a)}\left[\Gamma(a)\Gamma\left(a+\ff{2}{\beta}
\right)-\Gamma\left(a+\ff{1}{\beta}\right)^2\right]^{1/2},
\ee
one obtains
\bb \nn
W_a(\mu)&=&\ff{\beta}{\Gamma(a)}\Lambda^a \mu^{\beta a -1}
\exp\left(-\Lambda \mu^\beta\right),\\ \nn 
\Lambda&=&\Gamma(a)^{-\beta}\left[\Gamma(a)\Gamma\left(a+\ff{2}{\beta}\right)
-\Gamma\left(a+\ff{1}{\beta}\right)^2 \right]^{\beta/2}.
\ee
Thus here again, the distributions appearing in EVS are generalised
into distributions of sums, including a real parameter $a$ coming from
the asymptotic behaviour of the function $\Omega$.\\
Note that the generalised Fr\'echet and Weibull distributions found above
may also arise from a different statistical problem.
As already noticed by Gumbel \cite{gumbel} in the context of EVS,
the Fr\'echet and Weibull distributions may be related to the Gumbel
distribution through a simple change of variable. If a random variable
$X$ is distributed according to a Gumbel distribution $G_a$, then the
distribution of the variable $Y= e^{\lambda X}$ ($\lambda>0$) is a
generalised Fr\'echet distribution.
As $X$ is defined as a sum of correlated random variables associated with
a joint probability (\ref{JNgen}) with $P$ in the Gumbel class, $Y$ appears
as the product of variables with $P$ in the Fr\'echet class.
Similarly, the generalised Weibull distribution may also be
obtained from the Gumbel distribution by the change of variable
$Y=e^{-\lambda X}$.\\
In summary, one sees that the reformulation of standard EVS by means
of the joint probability (\ref{JNgen}) allows a natural and straightforward
definition of all the generalised extreme value statistics.
As already stressed, this generalisation breaks the equivalence with the
extreme value problem: there is no associated extreme process leading to the
generalised extreme value distributions $G_k$, $F_k$ or $W_k$ with $k$ real.
At the level of the joint probability (\ref{JNgen}), the equivalence
with EVS only holds if $\Omega(F)$ is a pure power-law of $F$,
with an \textit{integer} exponent $k$.
Yet, concerning the asymptotic distribution, extreme value distributions
are recovered as soon as $\Omega(F) \sim \Omega_0\, F^{k-1}$ for
$F \rightarrow 0$ (with $k$ integer), that is, when $\Omega$ is regular
in $F=0$.

\section{Physical interpretation}
Up to now we presented a mathematical result about sums of random variables,
linked by the joint probability (\ref{JNgen}). This joint probability leads
to non-Gaussian distributions that may be interpreted as the result of
correlations between those variables. All the informations about the
correlation are included in $J_N$. From a physicist point of view, this
is not completely satisfying: The degree of correlation is indeed usually
quantified by the correlation length. At first sight
it is disappointing that we can not extract such a quantity from
(\ref{JNgen}). One has to realise however that up to now we only dealt
with \textit{numbers}, without giving any physical meaning to any of those
numbers. To get some physical information from our result, we have to put
some physics in it first\footnote[1]{``Mathematician prepare abstract
reasoning ready to be used, if you have a set of axioms about the real world.
But the physicist has meaning to all his phrases.'' R.~P. Feynman in
\cite{Feynman}.}, giving an interpretation of the $u_n$, explaining also how
those quantities are arranged in time or space, introducing then a notion of
distance or time between the $u_n$ and the dimensionality of space.
The results presented in the previous sections could therefore describe
various physical situations. In this section we illustrate this idea by
studying a simple model.\\
Let us consider a one-dimensional lattice model with a continuous variable
$\phi_x$ on each site $x=1,...,L$. Although we do not specify the dynamics
explicitly, we have in mind that the $\phi_x$'s are strongly correlated.
One can define the Fourier modes
$\hat{\phi}_q$ associated to $\phi_x$ through
\bb \nn
\hat{\phi}_q = \sum_{x=1}^L \phi_x e^{iqx}.
\ee
A natural global observable is the integrated power spectrum (``energy'')
$E=\sum_x \phi_x^2$.
From the Parseval theorem, $E$ may also be expressed as a function of
$\hat{\phi}_q$ as $E = \sum_q |\hat{\phi}_q|^2$.
We now assume that the squared amplitudes $u_n=|\hat{\phi}_q|^2$, with 
$q=2\pi n/L$, follow the statistics defined by Eq.~(\ref{JNgen}).
For simplicity, we consider the simplest case where $P(z)=\kappa
e^{-\kappa z}$ and $\Omega(F)=F^{a-1}$, although more complicated
situations may be dealt with. This actually generalises the study
of the $1/f$ noise by Antal \etal \cite{antal01}.
The power spectrum is given by
\bb \label{moyphi}
\moy{|\hat{\phi}_q|^2} =
\left[ \kappa \left( \ff{|q|L}{2\pi}+a-1\right)\right]^{-1}
\propto \ff{1}{|q|+m},\\ \nn
m=\left(\ff{2\pi(a-1)}{L}\right)^{-1}.
\ee
The correlation function $C(r)$, given by the inverse Fourier transform
of $|\hat{\phi}_q|^2$,
\bb
C(r)=\mathcal{F}^{-1}\left[\moy{|\hat{\phi}_q|^2}\right](r),
\ee
can be computed using (\ref{moyphi}), leading to :
\bb \nn
C(r) \propto -\sin (mr)\; \mathrm{si}(mr) -\cos (mr)\;\mathrm{ci}(mr), 
\ee
where $\mathrm{si}$ and $\mathrm{ci}$ are respectively the sine and cosine
integral functions \cite{GR}. The 
correlation length is therefore defined as the typical length scale of $C(r)$:
\bb \nn
\xi=m^{-1}= \ff{L}{2\pi(a-1)}.
\ee
Thus $\xi$ appears to be proportional to the system size, which was
expected from the breaking of the central limit theorem
\footnote[2]{Dividing a system of linear size $L$ into (essentially)
independent subsystems with a size proportional to $\xi$, the number
of subsystems remains finite when $L \rightarrow \infty$ if $\xi \sim L$, 
so that the central limit theorem should not hold.}.
The particular case of a $1/f$ noise ($a=1$) then corresponds to a highly
correlated system, with $\xi/L \rightarrow \infty$. For $a \approx 1.5$,
a value often reported in physical systems \cite{bramwell00}, one gets
$\xi/L \simeq 0.15$: The correlation is weaker but $\xi$ still diverges
with $L$ \cite{clusel04}.\\
Altogether, this simple model may be thought of as a minimal model,
that allows some generic properties of more complex correlated physical
systems to be understood.\\
Obviously not all correlated systems will exhibit generalised EVS:
A well-known counter-example is the 2D Ising model at its critical
temperature \cite{Bruce81}.
Strictly speaking the generalised EVS can only be obtained
if it exists $P$ such as the joint probability is given by (\ref{JNgen}).
The number of observations of distributions close to a generalised Gumbe
 distribution suggests however that 
the expression (\ref{JNgen}) is general enough to reasonably approximate
the real joint probability in many situations. This would explain the
ubiquity of the generalised Gumbel in correlated systems.\\
More practically, the generalised extreme value distributions also appear
as natural fitting
functions for global fluctuations. If one is measuring fluctuations
of some global quantities, it seems quite reasonable to fit them with an
asymptotic distribution which could possibly take into account a violation
of the hypothesis of the central limit theorem. The generalised EVS are one
of the few such distributions. Using the reduced variable $\mu$, there is
only one free parameter in the Gumbel class, $a$, which quantifies the
deviation from the CLT.

\section{Conclusion}
In this article, we established that the so-called generalised extreme
value distributions are the asymptotic distributions
of random sums, for particular classes of random variables defined by
Eq.~(\ref{JNgen}), which do not
satisfy the hypothesis underlying the central limit theorem.
Interestingly, this is one of the very few known
generalisations of the central limit theorem to non-independent random variables. In this framework, it becomes clear that it is vain
to look for a hidden extreme process when one of the extreme value
distributions is observed in a problem of global fluctuations. Therefore
qualifying such distributions of generalised extreme value statistics is
somehow misleading.
The parameter $a$ quantifies the dependence of the random variables,
although further physical inputs are needed to give a physical interpretation
to this parameter. Within a simple model of independent Fourier modes,
$a$ is related in a simple way to the ratio of the correlation length
to the system size, a ratio that remains constant in the thermodynamic
limit for strongly correlated systems.
Besides, we believe that the classes of random variables
defined by Eq.~(\ref{JNgen}) may be regarded as reference classes
in the context of random sums breaking the central limit theorem, due to
their mathematically simple form.
Along this line of thought, it is not so surprising that distributions
close to the generalised Gumbel distribution are so often observed
in the large scale fluctuations of correlated systems, as for instance
in the case of the XY model at low temperature.

\appendix
\section{Fourier transform of the generalised Gumbel distribution}
In this Appendix we give an expression for the Fourier transform of
generalised Gumbel distribution $G_a$ defined by
\bb \nn
G_a(x)=\ff{\theta a^a}{\Gamma(a)} \exp \left[-a \theta (x+\nu)-a
e^{\theta (x+\nu)} \right],
\ee
with $\theta^2=\Psi'(a)$ and $\theta\nu=\ln a - \Psi(a)$.
The Fourier transform of $G_a$ is defined by
\bb \nn
\four{G_a}(\omega)&=&\int_{- \infty}^{+\infty} \dd x \; e^{-i\omega x} G_a(x).
\ee
Letting $u=a \exp(-\theta(x+\nu))$, we obtain
\bb \nn
\four{G_a}(\omega)&=&\ff{1}{\Gamma(a)}a^{-i\omega/\theta}e^{i\nu \omega}
\int_0^{\infty} \dd u \; u^{i\omega/\theta+a-1} e^{-u},\\ \nn
		&=&\ff{e^{i\nu \omega}}{a^{i \omega/\theta}}\ff{\Gamma
\left(\ff{i\omega}{\theta}+a \right)}{\Gamma(a)}.
\ee
Using the identity \cite{GR}
\bb \label{ID}
\Gamma(1+z)=e^{-\gamma z}\prod_{n=1}^{\infty}\ff{e^{z/n}}{1+\ff{z}{n}},
\ee
one finally obtains, with $\alpha=a-1$,
\bb \fl \nn
\four{G_a}(\omega)&=&\ff{e^{i\left(\nu-\ff{\gamma}{\theta}\right) \omega}}{
(1+\alpha)^{i\omega/\theta}} \prod_{n=1}^{\infty}\ff{\exp\left(\ff{i \omega}{
n\theta}\right)} {1+\ff{i \omega }{\theta(n+\alpha)}}\\ \fl \nn
&=&\left\{ \ff{e^{i\left(\nu-\ff{\gamma}{\theta}\right)\omega}}{(1+\alpha)^{
i \omega/\theta}} \exp \left[\ff{i \omega}{\theta} \sum_{n=1}^{\infty}
\left(\ff{1}{n}-\ff{1}{n+\alpha}\right) \right] \right\} \prod_{n=1}^{\infty}
\ff{\exp\left(\ff{i\omega}{n(\theta+\alpha)}\right)}
{1+\ff{i\omega}{\theta(n+\alpha)}}.
\ee

The identity (\ref{ID}) leads to
\bb  \label{r1}
\ff{\dd \ln \Gamma(1+\alpha)}{\dd \alpha}=-\gamma+\sum_{n=1}^{\infty}
\left(\ff{1}{n}-\ff{1}{n+\alpha}\right),
\ee
and therefore we obtain, using (\ref{sigma})
\bb \nn
\theta^2=\ff{\dd^2 \ln \Gamma(1+\alpha)}{\dd \alpha^2}=\sum_{n=1}^{\infty}
\ff{1}{(n+\alpha)^2}=\sigma^2.
\ee
Furthermore, the relation (\ref{r1}) leads to 
\bb \nn \fl
\left(\nu\theta-\gamma\right)-\ln\left(1+\alpha \right)+ \sum_{n=1}^{\infty}
\left(\ff{1}{n}-\ff{1}{n+\alpha}\right)&=&  (\nu \theta-\gamma)-\ln a+\left(\gamma
+\ff{\dd \ln \Gamma}{\dd a}\right),\\ \nn
&=&0.
\ee
So finally we have
\bb \nn
\four{G_a}(\omega)=\prod_{n=1}^{\infty}\ff{\exp\left(\ff{i\omega}
{\theta(n-1+a)}\right)} {1+\ff{i \omega}{\theta(n-1+a)}}.
\ee

\ack
It is a pleasure to thank Steven T. Bramwell, Michel Droz, Jean-Yves Fortin and Peter C.W. Holdsworth for useful discussions and comments. M.C. thanks University College London for hospitality during completion of this work.

\bibliographystyle{unsrt}

\end{document}